\documentclass[12pt]{article}
\usepackage{epsfig}

\usepackage{amssymb}
\usepackage{amsmath}
\usepackage{amsfonts}

\def\be{\begin{equation}}
\def\ee{\end{equation}}
\def\ba{\begin{array}{c}}
\def\ea{\end{array}}

\def\ben{$$}
\def\een{$$}

\newcommand{\bea}{\begin{eqnarray}}
\newcommand{\eea}{\end{eqnarray}}

\newcommand{\bbr}{\br\!\br}

\newcommand{\kt}{\rangle}
\newcommand{\br}{\langle}
\begin{document}

\titlepage

\vspace{.35cm}

 \begin{center}{\Large \bf

 Gegenbauer-solvable quantum chain model

  }\end{center}

\vspace{10mm}

 \begin{center}

 {\bf Miloslav Znojil}

 \vspace{3mm}
Nuclear Physics Institute ASCR,

250 68 \v{R}e\v{z}, Czech Republic

{e-mail: znojil@ujf.cas.cz}

\vspace{3mm}

%\today, kinemjpa.tex

\end{center}

\vspace{5mm}

%\newpage

\section*{Abstract}

$N-$level quantum model is proposed in which the energies are
represented by an $N-$plet of zeros of a suitable classical
orthogonal polynomial. The family of Gegenbauer polynomials
$G(n,a,x)$ is selected for illustrative purposes. The key novelty
lies in the use of non-Hermitian (a.k.a. cryptohermitan)
Hamiltonians $H\neq H^\dagger$. This  enables us to (1) start from
elementary secular equation $G(N,a,E_n)=0$, (2) keep our $H$, in the
nearest-neighbor-interaction spirit, tridiagonal, (3) render it
Hermitian in an {\em ad hoc}, non-unique Hilbert space endowed with
metric $\Theta\neq I$, (4) construct eligible metrics in closed
forms ordered by increasing nondiagonality and (5) interpret the
model as a smeared $N-$site lattice.

\newpage

 \section{Introduction \label{uvod} }

In atomic, molecular, nuclear and solid-state physics the simulation
of quantum phenomena via {\em finite-dimensional} Schr\"{o}dinger
equations
 \be
 H^{(N)}\,|\psi_n^{(N)}\kt=E_n^{(N)}\,|\psi_n^{(N)}\kt\,
 \label{SE}
 \ee
is often motivated numerically. Indeed, whenever a realistic
Hamiltonian gets approximated by its suitable $N$ by $N$
simplification $H=H^{(N)}$, the numerical solution of Eq.~(\ref{SE})
becomes routine \cite{Acton}, especially when our finite-dimensional
Hamiltonian is chosen tridiagonal,
 \be
 H^{(N)}=
  \left[ \begin {array}{ccccccc}
  %\hline
   a_0&c_0&0&0&\ldots&0&0
  \\
  %\hline
  b_1
&a_1&c_1&0&\ddots&&0
 \\0&b_2&a_2&c_2&\ddots&\ddots&\vdots
 \\0&0
 &\ddots&\ddots&\ddots&0&0
 \\
 %\hline
 {}\vdots&\ddots&\ddots
&b_{N-3}&a_{N-3}&c_{N-3}&0
 \\{}0&&\ddots&0&b_{N-2}&a_{N-2}&c_{N-2}
 \\{}0&0&\ldots&0&0&b_{N-1}&a_{N-1}\\
 %\hline
 \end {array} \right]\,.
 \label{kitie}
 \ee
At a fixed $N$, various $N$ by $N$ matrix problems (\ref{SE}) +
(\ref{kitie}) are often used in ambitious phenomenological
considerations since matrices $H^{(N)}$ (sometimes reinterpreted as
the so called chain-model Hamiltonians or lattice Hamiltonians with
the nearest-neighbor interaction) may mimic, say, a solid-state
spectrum of energies in condensed-matter physics. These concepts
found new applications in the context of apparently non-Hermitian
versions  (we would rather call them ``hiddenly Hermitian" or
``cryptohermitian" \cite{SIGMA} versions) of the XXZ spin chains
\cite{Korff}, of the Bose-Hubbard models \cite{Eva}, of the
Friedrichs-Fano-Anderson tight-binding lattice models \cite{fano},
of the tightly bound lattices of electrons \cite{Jin}, optical
lattices \cite{Jinbe} etc. There exist many other papers which are
also certainly worth mentioning. In their incomplete complementary
sample we would like to attract attention of the reader to the close
connections between non-Hermitian chain models and the so called
Reggeon field theory \cite{A1} or to Ising model and quantum spin
chains \cite{A2}.

The combined mathematical and physical appeal of the generic
discrete and tridiagonal models (\ref{kitie}) seems partially marred
by the more or less purely numerical \cite{wilk} or perturbative
\cite{pert} nature of their solution. For this reason, analytically
solvable models are often preferred in analysis \cite{PRB}. Some
authors simplified mathematics by paying attention to the effects
connected with the restricted, one-parametric variation of the
end-site-interaction matrix elements $c_j$ and $b_{j+1}$ with $j=0$
and $j=N-2$ in Eq.~(\ref{kitie}) \cite{Jin,david}.

Marginally we could add that similar discrete solvable models with
pairs of point-like interactions played important role in the recent
extensive discussion of some conceptual problems of cryptohermitian
quantum scattering \cite{Jones,Jonesscattsmeared}. In this context a
lot of misunderstandings emerged when people forgot to distinguish
between the ``formal coordinate" $x$ (often chosen as playing the
role of the argument in wave functions $\psi(x)$) and the
``observable coordinate" (a position-operator eigenvalue denoted by
another symbol, say, $q$). In a very well written paper~\cite{Hoo}
interested reader may find the nice explanation of this subtlety
emerging as highly relevant even on the very elementary level of
mathematics used in introductory textbooks on quantum mechanics.

Once one moves to the more sophisticated cryptohermitian models
where the ``formal coordinate" $x$ itself ceases to be observable,
the concept of ``locality" must be reconsidered and used with
enhanced care. For example, a very instructive comment given in
section 5 of Ref.~\cite{cubic} shows that the formal wave function
of a physical {\em localized} state may look non-local as a function
$\psi(x)$ of the formal coordinate $x$.

In opposite direction it has been noticed and emphasized by Jones
\cite{Jones} that in virtually any experimentally oriented setup we
usually treat interaction $V$ as if it were {\em prepared} as a
specific function of the measurable coordinate $q$. In this sense,
the crucial role of the specification of observables and of the
difference between $x$ and $q$ gets even more important in
non-Hermitian setting \cite{Geyer}.

In order to circumvent similar complications a number of papers
studied just bound-state problems and preferred their exactly
solvable non-Hermitian models \cite{A3}. The
solvability-guaranteeing simplifications may reduce the menu of
interesting phenomena. Typically, the simplified models explain the
emergence of fragile, unstable components in the spectra \cite{robu}
but they can hardly compete with realistic models in offering
sufficient variability of the parametric dependence of the
energies~\cite{Eva}. The spectra obtained in the simplified solvable
model of Ref.~\cite{Jin} admit, for example, just a very special
form of the confluence of energy pairs while a much richer menu of
quantum catastrophes of this category may exist in general~\cite{A}.

A remedy has been found in Ref.~\cite{D}. We revealed that there
exist {\em non-numerical} chain models or quantum lattices
(\ref{kitie}) with a much less restricted qualitative variability of
spectra. These models were characterized by a delocalized
interaction exhibiting an up-down symmetry. The pairs of sites with
indices $m$ and $N-m$ were attached the {\em same} strength of
impurity or interaction. Although the productivity of such an
artificial assumption was reconfirmed, say, in refs.~\cite{Jin} and
\cite{E}, its physical interpretation remained obscure. One would
like to have some exactly solvable quantum-lattice models {\em
without} such a symmetry. This motivated our present analysis during
which we developed {another} class of solvable quantum-lattice
models of form (\ref{kitie}) {without} similar non-local, long-range
auxiliary correlation.

\section{Gegenbauer-polynomial quantum lattice\label{prvacka}}

In connection with the definition of the concept of solvability
misunderstandings frequently emerge. The puzzle may find different
resolutions. In a context-dependent way the property of being
solvable is assigned, e.g., to differential Hamiltonians $H =
p^2+V(x)$ for which {\em all} of the wave functions $\br x
|\psi_n\kt$ of bound states prove proportional to suitable classical
orthogonal polynomials~\cite{Levaiold}. In our present paper we
shall transfer such a definition of exact solvability to the
difference and finite-matrix equations. Thus, we shall {\em
postulate} that the $N-$plet of our $N-$dimensional bound-state
vectors $|\psi_n\kt$ in Eq.~(\ref{SE}) is given in advance.

Naturally, the most straightforward definition of these vectors
would specify them directly in terms of some classical orthogonal
polynomials. For the sake of brevity we shall solely pay attention
to Gegenbauer polynomials $G(n,a,x)$ ($=C^a_n(x)$ in \cite{Ryzhik}
or $C^{(a)}_n(x)$ in \cite{Stegun}; our notation is taken from MAPLE
\cite{Maple}). As long as these (sometimes called ultraspherical)
polynomials degenerate to the different (viz., Chebyshev)
polynomials at $a = 0$, we shall assume that $a >0$. In this case
they satisfy the well known recurrence relations
 \be
      n\,  G(n,a,x) = 2\,(n+a-1)\,x\,G(n-1,a,x)
                       - (n+2\,a-2)\,G(n-2,a,x)
                       \label{reku}
                       \ee
at $n =1, 2, \ldots$, with initial  $G(0,a,x) = 1$ and $G(1,a,x) =
2\,a\,x$.

In the initial step of our constructive considerations we shall
guarantee the validity of our above-mentioned matrix Schr\"{o}dinger
Eq.~(\ref{SE}) by assuming its formal coincidence with the truncated
version of recurrences~(\ref{reku}). This means that we shall just
use the following input form of the bound-state eigenvector,
 \be
 |\psi_n^{(N)}\kt=\left (
  \ba
 \br 0|\psi_n^{(N)}\kt=G(0,a,E_n)\\
 \br 1|\psi_n^{(N)}\kt=G(1,a,E_n)\\
 \vdots\\
 \br N-1|\psi_n^{(N)}\kt=G(N-1,a,E_n)
 \ea
 \right )\,
 \label{input}
 \ee
and determine the $n-$th energy level $E_n$ as the value of
coordinate $x$ at which recurrences~(\ref{reku}) terminate. Thus,
every energy will coincide with one of the roots of the closed-form
secular equation
 \be
 G(N,a,E_n)=0\,.
 \label{secu}
 \ee
Our Gegenbauerian Hamiltonian $H=H(a)$ will just mimic  recurrences
(\ref{reku}). Its main diagonal will vanish (i.e., we set
$a_0=a_1=\ldots=0$ in (\ref{kitie})) and  the pair of non-vanishing
neighboring diagonals will be composed of elements numbered by $j =
0, 1, \ldots, N-2$,
 \be
  c_j=c_j(a)=1/(2a+2j)\,,\ \ \ \
 b_{j+1}=b_{j+1}(a)=(2a+j)/(2a+2j+2)\,.
 \label{geg}
 \ee
This idea forms the starting point of our abstract message in its
concrete Gegenbauer-polynomial realization. Within the more general
class of quantum lattices and discrete models (\ref{kitie})
exemplified by such a choice the matrix elements are real but the
matrix $H$ itself is, generically, asymmetric, i.e., non-Hermitian.
Fortunately, its spectrum is real (i.e., potentially observable) so
that we are allowed to treat this $H$ as an {\em exactly solvable}
effective Hamiltonian of a quantum system with the prescribed
segment of spectrum fitted by an $N-$plet $E_n^{(N)}(a)$ of roots of
Gegenbauer polynomial $G(N,a,E)$.

\section{(Hidden) Hermiticity }

It is known that the manifest non-Hermiticity feature does not
disqualify operator $H\neq H^\dagger$  from being used as a
Hamiltonian of a quantum system. After all, not too dissimilar
non-Hermitian phenomenological Hamiltonians (complex and acting in a
finite-dimensional vector space) were used in Refs. \cite{Korff} -
\cite{Jinbe}. Interested reader may find a compact introduction into
quantum theory with similar cryptohermitian Hamiltonians either in
our review~\cite{SIGMA} or in this section.

In essence, we must get rid of the overrestrictive and most
elementary (often called ``Dirac's" \cite{Carl}) requirement of the
current but very special Hermiticity defined via the mere vector or
matrix transposition accompanied by complex conjugation. This
defines dual vectors called, in the conventional textbook language,
``Dirac's bra-vectors",
 \be
 {\cal T}^{(Dirac)}: \ |\psi \kt \ \to \
   \br \psi|\,.
 \ee
The choice of ${\cal T}^{(Dirac)}$ (represented just by appended
superscript $^\dagger$ when applied to operators) is {\em not} the
only option. In models with Dirac-non-Hermiticity $H\neq H^\dagger$
we must {\em necessarily} use {another}, less trivial definition of
Hermitian conjugation. The point is that after such a change of
definition our operator $H$ may become self-adjoint and compatible
with postulates of Quantum Mechanics.

The transition to general Hermitian conjugation will require a
modification of conventional notation. Firstly, the ``new" dual
vectors must be defined by  generalized formula
 \be
 {\cal T}^{(\Theta)}: \ | \psi \kt \ \to \
   \bbr \psi| := \br \psi |\,\Theta\,
   \label{conjuga}
 \ee
where matrix $\Theta$  is called ``metric" \cite{Geyer} and where,
whenever $\Theta\neq I$, the resulting dual vectors are marked as
``brabras". Secondly, the same danger of misunderstanding threatens
the application of the non-Dirac Hermitian conjugation to operators
${\cal A}$ so that we recommend it to be marked by a different
(viz., doubled) superscript,
 \be
  {\cal A} \ \to \  {\cal A}^\ddagger :=\,\Theta^{-1}\,
  {\cal A}^\dagger\,\Theta\,.
  \label{conju}
 \ee
In the spirit of any good textbook on Linear Algebra, Functional
Analysis or Quantum Mechanics the metric must be required
invertible, Hermitian and positive definite \cite{Geyer}. After two
notation innovations (\ref{conjuga}) and (\ref{conju}) the formalism
of Quantum Theory remains unchanged. On the level of notation the
symbol of double bras ($\bbr$) will replace all the Dirac's simple
bras ($\br$), especially whenever a mean value, physical probability
or measurements are concerned. Similarly, in formulae carrying
physical meaning the simple superscripts $^\dagger$  must be all
replaced by their doubled forms $^\ddagger$. The ``false"
representation ${\cal H}^{(F)}$ of the Hilbert space with the
Dirac's unacceptable $\Theta^{(F)}=I$ must {\em consequently} be
replaced by the ``standard" Hilbert space ${\cal H}^{(S)}$ of
physical states $\psi$.

\section{Hilbert-space metrics}
%$\Theta=\Theta(H)$ }

From the pragmatic point of view the theoretical imperatives of
preceding section may be softened, during practical calculations, by
staying in the naive (and, by assumption, much friendlier) Hilbert
space ${\cal H}^{(F)}$ and by the treatment of the obligatory
doubled bras $\bbr$ of Eq.~(\ref{conjuga}) and doubled superscripts
$^\ddagger$ of Eq.~(\ref{conju}) as mere abbreviations. In
Ref.~\cite{SIGMA} we summarized further reasons for a {\em parallel}
use of spaces ${\cal H}^{(S)}$ together with their ``friendly -
false" partners ${\cal H}^{(F)}$. Firstly, just the knowledge of the
matrix $\Theta$ (which must be self-adjoint in ${\cal H}^{(F)}$
\cite{Geyer}) is fully sufficient for all purposes. Secondly, the
key guarantee of unitarity of the evolution generated by $H$ in
${\cal H}^{(S)}$ (where $H = H^\ddagger$ and $\Theta \neq I$) gets
easily legible in ${\cal H}^{(F)}$ via ``translations"
(\ref{conjuga}) and (\ref{conju}). In fact, a deeper explanation of
this point deserves the (following) separate paragraph.

\subsection{Dieudonn\'{e} equation }

In Ref.~\cite{SIGMA} we explained the way in which the Hermiticity
of $H$ in ${\cal H}^{(S)}$ (based on the nontriviality of metric
$\Theta^{(S)}\neq I$) may be understood as equivalent to the
manifest Hermiticity of a suitable isospectral operator
 \be
 \mathfrak{h}=\Omega\,H\,\Omega^{-1}=\mathfrak{h}^\dagger\,.
 \label{ouzo}
 \ee
The latter operator is defined, in principle, in another, third
Hilbert space ${\cal H}^{(P)}$ with trivial metric $\Theta^{(P)}=I$
(the superscript stands for ``paternal" or ``physical"). It is
assumed that spaces ${\cal H}^{(P)}$ and ${\cal H}^{(S)}$ are
unitary equivalent so that we may recall Eq.~(\ref{conju}), deduce
 \be
 \mathfrak{h}^\dagger=\left (\Omega^{-1}
 \right )^\dagger\,H^\dagger\,\Omega^\dagger\,,
 \label{spouzo}
 \ee
abbreviate $ \Omega^\dagger \Omega:=\Theta$ and end up with the
relation
 \be
 H^\dagger\,\Theta=\Theta\,H\,
 \label{dieudo}
 \ee
dating back to the old paper by Dieudonn\'e \cite{Dieudonne}. That's
why we shall call Eq.~(\ref{dieudo}) ``Dieudonn\'e's equation" in
what follows, keeping in mind that this is meant in a loose sense
since Dieudonn\'e himself admitted that $\Theta$ in (\ref{dieudo})
might not be invertible.

For our finite-dimensional real Hamiltonians $H=H^{(N)}$ which are
given in advance, the latter equation forms the set of $N^2$
constraints imposed upon the $[N(N+1)/2]-$plet of the unknown real
matrix elements of matrix $\Theta=\Theta^\dagger$. Our task may now
be formulated as a non-numerical construction of complete solution
of this linear algebraic system.

\subsection{The method of  solution }

The constructive way of making Hamiltonian $H$ and metric $\Theta$
compatible with Dieudonn\'e's Eq.~(\ref{dieudo}) is not too easy in
general. The main result of our paper will be the {non-numerical}
construction of the {\em general} metric $\Theta$ which will satisfy
Eq.~(\ref{dieudo}) for the Gegenbauerian input Hamiltonian
$H^{(N)}(a)$. {\it Ipso facto}, this will also make our Hamiltonian
self-adjoint in the respective physical Hilbert space ${\cal
H}^{(S)}$.

In full detail, the construction of metrics will be described in
section \ref{hla} below. In a preparatory phase let us now just
explain its key ideas. Firstly, in the light of the linearity of
Eq.~(\ref{dieudo}) we shall assume that the metric may be sought in
the form of superposition of certain simpler matrices ${\cal P}$
which will satisfy the same equation,
 \be
 \left (H^{(N)}(a)\right )^\dagger\,{\cal P}
 ={\cal P}\,H^{(N)}(a)\,,
 \label{dieudonear}
 \ee
but which will not necessarily be invertible or positive definite.
Secondly, we shall assume that these ``pseudometric" matrices will
form an $N-$plet of linearly independent solutions ${\cal P}={\cal
P}_k^{(N)}(a)$ with $k=0,1,\ldots,N-1$. This will enable us to
search for the metric in the form
 \be
 \Theta=\Theta^{(N)}(\vec{\alpha},a)= \alpha_0\,\Theta_0^{(N)}(a)
 + \sum_{k=1}^{N-1}\,\alpha_k\,{\cal P}_k^{(N)}(a)\,
 \label{metrika}
 \ee
where the variability of the $N-$plet of real parameters
$\vec{\alpha}=(\alpha_0,\alpha_1,\ldots,\alpha_{N-1})$ will only be
restricted by the requirement of the positivity of the matrix
$\Theta^{(N)}(\vec{\alpha},a)$.

The concrete implementation of the requirement of the simplicity of
the individual auxiliary pseudometrics ${\cal P}_k$ is
model-dependent. For our present model their explicit construction
proved feasible when we assumed that every ${\cal P}_k$ is a
$(2k+1)-$diagonal matrix. This assumption itself resulted from the
experience which we gained during the similar constructions of
metrics as performed in Ref.~\cite{fund}. This experience also
facilitated the organization of our concrete recurrent calculations.

The key idea of our present non-numerical algorithm of solution of
Eq.~(\ref{dieudonear}) remained the same as in Ref.~\cite{fund}. In
concrete applications we shall see how this recipe employs the
chess-board-like ``coloring" of elements of relevant matrices. In
this manner, each Hamiltonian $H$ gets separated into its
``white-field matrix elements" (say, all elements $H_{j,k}$ with
$|j-k|=$even) and ``black-field matrix elements" (i.e., elements
$H_{j,k}$ with $|j-k|=$odd). Once the same coloring is applied to
the ansatz for the metric $\Theta$ (or rather to each indefinite and
sparse pseudometric ${\cal P}_k^{(N)}(a)$), one is immediately able
to decompose Eqs.~(\ref{dieudo}) and/or (\ref{dieudonear}) into
their ``same-color" subsystems and to develop and employ some
suitable ansatzs for their recurrent solution.

One should not forget that even before finishing the systematic
construction of {\em all} of the components ${\cal P}_k^{(N)}(a)$ of
the metric we may interrupt the process and turn attention to the
truncated versions of series (\ref{metrika}),
 \be
 \Theta_k^{(N)}(\vec{\alpha}',a)=  \Theta_0^{(N)}(a)
 + \sum_{j=1}^{k}\,\alpha_j\,{\cal P}_j^{(N)}(a)\,.
 \label{metrikabe}
 \ee
Here, the mere $k$ free parameters $\alpha_j$ appear arranged in a
shorter, primed array $\vec{\alpha}'$. One should also pay attention
to the fact that in Ref.~\cite{fund} as well as in our present model
the $k-$subscripted special metrics (\ref{metrikabe}) still remain
sparse, containing just $2k+1$ non-vanishing diagonals. The latter
observation will certainly facilitate our ultimate task of imposing
the positivity requirements upon expansions (\ref{metrika}) or
(\ref{metrikabe}) of the metric.

\section{Diagonal metrics }

\subsection{The construction of $\Theta_0(a)$ }

All the details of the implementation of our above recipe depend on
the form of the input Hamiltonian $H$. For its Gegenbauerian choice
given by Eq.~(\ref{geg}), this Hamiltonian is an extremely
elementary, purely ``black-field" matrix, rendering the recurrent
solution of Eq.~(\ref{dieudo}) particularly straightforward. For
illustration purposes let us now consider the diagonal (i.e., $k=0$)
ansatz
 \be
 \Theta_0(a)=
 \left[ \begin {array}{ccccc}
 \theta_0&0&\ldots&0&0\\
 0&\theta_1&0&\ldots&0\\
 \vdots&\ddots&\ddots&\ddots&\vdots\\
 0&\ldots&0&\theta_{N-2}&0\\
 0&0&\ldots&0&\theta_{N-1}
 \end {array} \right]\,.
 \label{ans0}
 \ee
As long as the individual matrix elements will {\em not} vary with
the growth of dimension $N$, we may leave the value of $N$
unspecified. The inspection of recurrences (\ref{dieudo}) then
reveals that they connect just equal-color elements. This means that
{\it a priori}, ansatz (\ref{ans0}) may lead to nontrivial
solutions. We may start their recurrent construction from any
nonvanishing element, say, from $\theta_0=2a^2$. After a
comparatively tedious algebra this choice of normalization leads to
the compact and transparent final result with $\theta_1=a+1$ and
with
 \be
 \theta_j=\frac{a+j}{(1+2a)(2+2a)\ldots(j-1+2a)}\
 \label{expliciti}
 \ee
at all the remaining $ j=2,3,\ldots,N-1$.

\subsection{A comment on matrices $\mathfrak{h}$}

An important feature of the above-constructed metric $\Theta_0(a)$
is that it is easily invertible and manifestly positive definite at
any $a>0$ and at any $N\geq 1$. The existence of such a metric is an
important merit of the model because we may now recall relation
(\ref{ouzo}), define the matrix elements of the simplest auxiliary
matrix $\Omega=\Omega_0$,
 \be
 \left (\Omega_0 \right )_{mn} = \delta_{mn}\, \sqrt{\theta_n}\,
 \label{nizoze}
 \ee
and obtain finally the simplest explicit partner Hamiltonian
 \be
 \mathfrak{h}_0^{(N)}(a)=
 \left[ \begin {array}{cccccc}
 0&\mu_0&0&0&\ldots&0\\
 \mu_0&0&\mu_1&0&\ddots&\vdots\\
 0&\mu_1&0&\mu_2&\ddots&0\\
 \vdots&\ddots&\ddots&\ddots&\ddots&0\\
 0&\ldots&0&\mu_{N-3}&0&\mu_{N-2}\\
 0&\ldots&0&0&\mu_{N-2}&0
 \end {array} \right]\,
 \label{ans000}
 \ee
acting in space ${\cal H}^{(P)}={\cal H}^{(P)}_0$, isospectral with
our original non-Hermitian matrix $H^{(N)}(a)$ and possessing matrix
elements easily derived in closed form,
 \be
 \mu_k=\frac{1}{2}\,\sqrt{\frac{2a+k}{(a+k)\,(a+k+1)}}\,,
 \ \ \ \ \ k = 0, 1, \ldots ,N-2.
 \ee
Due to the unitary equivalence between Hilbert spaces ${\cal
H}^{(P)}_0$ and ${\cal H}^{(S)}={\cal H}^{(S)}_0$ we may conclude
that Eq.~(\ref{ans0}) represents the simplest possible Hermitization
of our Gegenbauer-oscillator Hamiltonian $H^{(N)}$.

The existence of the partner Hamiltonian (\ref{ans000}) trivially
re-confirms the well known fact that the spectra of energies
$E_n^{(N)}(a)\, $ defined by Eq.~(\ref{secu}) are all
real~\cite{Stegun}. Moreover, the manifest positivity and
diagonality of $\Theta_0(a)$ makes the explicit construction of
matrix $\Omega_0$ virtually trivial. The latter observation is not
easily transferred to other models. For example, interested readers
may consult Ref.~\cite{cubic} showing that and in which way a very
simple Hamiltonian $H$ may be assigned extremely complicated
isospectral partners $\mathfrak{h}$.

Exceptions from the latter generic rule exist. In the present
context of models on lattices a typical one has been found in
paper~\cite{example}. A non-diagonal, band-matrix metrics $\Theta$
has been shown there to admit a transparent, sparse-matrix structure
of factors in $\Theta=\Omega^\dagger \Omega$ as well as of the
corresponding isospectral Hamiltonian $\mathfrak{h}$. Of course,
this type of result must be considered exceptional. Formally, the
reason is that  the use of formula~(\ref{ouzo}) which defines the
partner Hamiltonian $\mathfrak{h}$ requires the explicit knowledge
of the inverse matrix $\Omega^{-1}$ which is usually not a sparse
matrix even if $\Omega$ itself is.

This being said it is necessary to admit that one cannot exclude
that our present Gegenbauerian example will prove exceptional and
that it will also admit the existence of compact formulae for
$\mathfrak{h}$, e.g., at some coordinate-smearing choice of
$k=k_{(exceptional)}\geq 1$. With the notable exception of our
knowledge of tridiagonal $k_{(exceptional)}=0$ matrix~(\ref{ans000})
the existence and possible structure of such formulae is an open
problem at present. In fact, the lack of our explicit knowledge of
all of the manifestly Hermitian Hamiltonians $\mathfrak{h}$ hinders,
first of all, the most common strategy of interpretation of the
system in question illustrated, e.g., in Ref.~\cite{Batal} and based
on the correspondence principle applied directly inside ${\cal
H}^{(P)}$.

A positive aspect of the existence of missing parts of the puzzle is
that if any relevant matrix $\mathfrak{h}$ really remained
sufficiently simple and defined in closed and compact form, all the
reasons for working with its equivalent representation $H$ in ${\cal
H}^{(S)}$ would in fact be lost. The situation is similar to the
preference of $H$ in nuclear physics \cite{Geyer} where the more
complicated partner $\mathfrak{h}$ is even well known in advance.
The same preference of the maximally simple representation of the
Hamiltonian remains recommended for concrete calculations even
though we proceed here in opposite direction, viz, from the choice
of $H$ to the construction of its Hermitizations mediated by
$\Theta$s in alternative Hilbert spaces ${\cal H}^{(S)}$.

\section{Band-matrix metrics  \label{hla} }

It has been explained in Refs.~\cite{Jonesscattsmeared} and
\cite{fund} that tridiagonal metrics, i.e., in our case, the
one-parametric family of matrices
 \be
 \Theta_1^{(N)}(\alpha_1,a)=  \Theta_0^{(N)}(a)
 + \alpha_1\,{\cal P}_1^{(N)}(a)\,
 \label{metrikabece}
 \ee
simulate a nearest-neighbor smearing of coordinates while the
pentadiagonal metrics
 \be
 \Theta_2^{(N)}(\alpha_1,\alpha_2,a)=  \Theta_0^{(N)}(a)
 + \alpha_1\,{\cal P}_1^{(N)}(a)
 + \alpha_2\,{\cal P}_2^{(N)}(a)\,
 \label{metrikabeceda}
 \ee
may mimic a next-to-nearest neighborhood smearing, etc. In this
manner the index $k$ in Eq.~(\ref{metrikabe}) is tractable as a
certain measure of a {\em dynamical}, Hilbert-space-related
``nonlocality" of the quantized lattices in question.

\subsection{Tridiagonal metrics $\Theta_1^{(N)}(\alpha_1,a)$}

In Gegenbauer example~(\ref{geg}) all the generalized $k=1,2,\ldots$
metrics (\ref{metrika})  may be constructed in closed form,
non-numerically, by the recurrent solution of Eq.~(\ref{dieudo}).
After some trial-and-error experimenting the first nontrivial,
tridiagonal metric $\Theta_1^{(N)}(\alpha_1,a)$ (containing just the
single item in the primed array of parameters $\vec{\alpha}'\equiv
\alpha_1$) may be found via the tridiagonal (or, more strictly
speaking, bidiagonal) $k=1$ ansatz for its only nontrivial
sparse-matrix component
 \be
 {\cal P}_1^{(N)}(a)=
 \left[ \begin {array}{cccccc}
 0&\kappa_1&0&\ldots&0&0\\
 \kappa_1&0&\kappa_2&0&\ldots&0\\
 0&\kappa_2&0&\kappa_3&\ddots&\vdots\\
 \vdots&\ddots&\ddots&\ddots&\ddots&0\\
 0&\ldots&0&\kappa_{N-2}&0&\kappa_{N-1}\\
 0&0&\ldots&0&\kappa_{N-1}&0
 \end {array} \right]\,.
 \label{ans1}
 \ee
The combined use of the experience and computer algebra leads to the
truncation-independent result. Using the convenient initial
$\kappa_1=2a$ and  $\kappa_2=1$ one obtains the closed formula
 \be
 \kappa_j=\frac{1}{(1+2a)(2+2a)\ldots(j-2+2a)}\
 \ee
for the solution (\ref{ans1}) of Eq.~(\ref{dieudo}) valid at all $
j=3,4,\ldots,N-1$. Let us re-emphasize that these matrix elements
exhibit the remarkable property of not changing their form with the
matrix dimension~$N$.

\subsection{The domains of positivity of metrics $\Theta_1^{(N)}(\alpha_1,a)$}

\begin{figure}[h]                     %instead of \begin{figure}[t]
\begin{center}                         %instead of \begin{center}
\epsfig{file=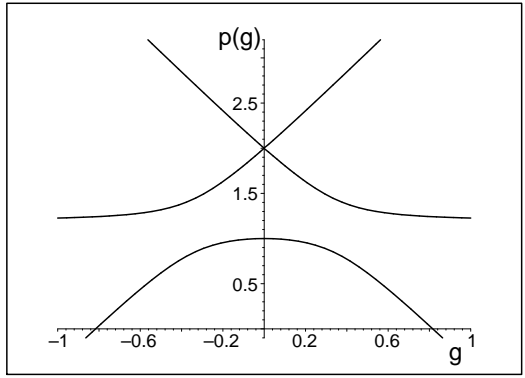,angle=270,width=0.5\textwidth}
\end{center}                         %instead of \end{center}
\vspace{-2mm}\caption{Three eigenvalues $p=p(g)$ of metric
$\Theta_1^{(3)}(g,1)$.
 \label{fied}}
\end{figure}

It is worth noticing that the positive definiteness of the
tridiagonal metrics (\ref{metrikabece}) would be lost for larger
$\alpha_1>\alpha_{critical}^{(N)}(a)$. Using an analytic method this
expectation may be illustrated via a slightly renormalized
two-dimensional metric
 $$
 \Theta_1^{(2)}(b/2,a)=
\left[ \begin {array}{cc}
2\,{a}^{2}&ab\\\noalign{\medskip}ab&a+1\end {array} \right]
 $$
possessing two real eigenvalues
 $$
 1/2\,a+1/2+{a}^{2}\pm 1/2\,\sqrt {-3\,{a}^{2}+2\,a-4\,{a}^{3}+1+4\,{a}^{4
}+4\,({ab})^{2}}\,.
 $$
It is easy to deduce that the domain $D$ of positivity of this
metric coincides with the interval of
 $$
 b \in (-\sqrt {2\,a+2},\sqrt {2\,a+2})\,.
  $$
At $N>2$ a graphical determination of the domains $D^{(N)}$ may be
used. For illustration let us consider $N=3$ and metric
 $$
 \Theta_1^{(3)}(\alpha_1,a)=
 \left[ \begin {array}{ccc} 2\,{a}^{2}&2\,\alpha_1a&0\\\noalign{\medskip}2\,g
a&a+1&\alpha_1\\\noalign{\medskip}0&\alpha_1&{\frac
{a+2}{2\,a+1}}\end {array}
 \right]
 $$
with the $\alpha_1-$dependence of its three eigenvalues illustrated
by Figure~\ref{fied} at $a=1$.

%********** Figure 1 zde

The pattern of the graphical localization of the eigenvalues of our
tridiagonal metrics $\Theta_1^{(N)}(\alpha_1,a)$ remains
qualitatively very similar in a broad range of parameters $N$,
$\alpha_1$ and $a$. In particular, we may be sure that the matrix
$\Theta_1^{(N)}(\alpha_1,a)$ remains positively definite at all the
sufficiently small nondiagonalities, i.e., in a nonempty subdomain
of $D^{(N)}$ where $|\alpha_1| \ll a$.

Several interesting as well as practically highly relevant questions
arise when one tries to extend the graphical analysis to higher
dimensions $N$. First of all, the growth of the necessary numerical
precision makes the analysis a bit costly. Indeed, one must be
careful with the numerical localization of the eigenvalues of the
metric because even our fully explicit formula (\ref{expliciti})
leads to a perceivable numerical contrast between the maximal
eigenvalues $\theta_0=\theta_1=1$ and the unexpectedly quickly
decreasing roots $\theta_7\sim 0.0003968$ or $\theta_8\sim
0.0000496$ (etc) of the corresponding secular equation.

Fortunately, the extremely elementary form of the matrix elements of
our Gegenbauerian tridiagonal metrics (\ref{metrikabece}) still
supports the practical feasibility of the direct numerical
localization of the boundaries of the related two-dimensional
domains $D_1^{(N)}$ of admissible parameters $\alpha_1$ and $a$ up
to the fairly large dimensions. Moreover, there exists an
encouraging numerical evidence that these boundaries $\partial
D_1^{(N)}$ stabilize and remain only very weakly dependent on the
dimension at  large $N \gg 1$.

\begin{table}[t]
\caption{The $N-$dependence of boundaries $\pm G$ of the domain
$D_1^{(N)}$ at $a=1$.} \label{pexp4}

\vspace{2mm}

\centering
\begin{tabular}{||c|l|l|l||}
\hline \hline
  $N$ & {\rm boundary value} $G$&{\rm neighboring $G'$ }&{\rm next $G''$ }\\
 %\multicolumn{1}{c|}{$\vdots$}\\
 \hline \hline
 1&$\infty$& --- & ---
 \\
 2&$1$& --- & ---
 \\
 3&$0.8164965809$& --- & ---
 \\
 4&$0.7835809235$& 2.210430034 & ---
 \\
 5&$0.7772152453$& 1.528761895 & ---
 \\
 6&$0.7761738933$& 1.347821298 & 3.702152325
 \\
 7&$0.7760367842$& 1.284679682 & 2.333798009
 \\
 8&$0.7760220038$& 1.261982266 & 1.922171587
 \\
 9&$0.7760206592$& 1.254396565 & 1.747726425
 \\
 \hline \hline
\end{tabular}
\end{table}

%
%plot([[1/2,1],[1/3,0.8164965809],[1/4,0.7835809235],[1/5,0.7772152453],
%[1/6,0.7761738933],[1/7,0.7760367842],[1/8,0.7760220038],[1/9,0.7760206592]]);

A persuasive sample of such an evidence is provided  by Table~1
where we choose $a=1$ and tabulated the values of $G$ (our metric
$\Theta_1^{(N)}(g,1)$ is positive definite for $g \in (-G,G)$)
together with auxiliary values $G'$ and $G''$  (our metric
$\Theta_1^{(N)}(g,1)$ has at most one or at most two negative
eigenvalues in the larger intervals $g \in (-G',G')$ and  $g \in
(-G'',G'')$, respectively).

The Table strongly and very persuasively supports the
$N-$independence of $G\approx 0.776$ in the limit $N \to \infty$
(i.e., the existence and stability of a non-empty domain $D_1^{(N)}$
where the metric is positive). Indeed, the left column of the Table
indicates that the $N-$th value of $G$ only differs from its
predecessor in the $(N-3)-$rd decimal digit.

\subsection{Pentadiagonal metrics}

At $k=2$ and variable $N$ we may try to solve Eq.~(\ref{dieudo}) by
pentadiagonal ansatz
 \be
 {\cal P}_2^{(N)}(a)=
 \left[ \begin {array}{cccccccc}
 0&0&\gamma_1&0&0&0&\ldots&0\\
 0&\delta_1&0&\gamma_2&0&0&\ldots&0\\
 \gamma_1&0&\delta_2&0&\gamma_3&0&\ddots&\vdots\\
 0&\gamma_2&0&\delta_3&0&\gamma_4&\ddots&0\\
 0&0&\ddots&\ddots&\ddots&\ddots&\ddots&0\\
 \vdots&\ddots&\ddots&\gamma_{N-4}&0&\delta_{N-3}&0&\gamma_{N-2}\\
 0&\ldots&0&0&\gamma_{N-3}&0&\delta_{N-2}&0\\
 0&0&\ldots&0&0&\gamma_{N-2}&0&\omega^{(N)}
 \end {array} \right]\,
 \label{ans2}
 \ee
using the same recurrent method as above. The selection of
$\gamma_1=a$ and the consistent specification of
$\gamma_2=(1+a)/(4+2a)$ initiate now the combined recurrences for
two unknown sequences in (\ref{ans2}). These recurrences may be
extracted, from linear algebraic Eq.~(\ref{dieudo}), as a subset of
all of its linearly independent items. The result of their solution
(which was, naturally, computer-assisted and rather lengthy) can be
written down in closed form, with $j=3,4,\ldots,N-1$ in
 \be
 \gamma_j=
 \frac{1+a}
 {(2j+2a)\Gamma_{j-2}}\,,\ \ \ \ \Gamma_{n}=(1+2a)(2+2a)\ldots(n+2a)\,
 \ee
and with $\Gamma_{0}=1$ and $j=1,2,\ldots,N-1$ in
 \be
 \delta_j=
 \frac{2[2a^3+3a^2-(4j-5)j\,a-(2j^2-1)(j-1)]}
 {(2j+2+2a)(2j-2+2a)\Gamma_{j-1}}
  \,.
  \label{lattere}
 \ee
At the smallest subscripts $j$  there occur incidental
factorizations  which simplify slightly the numerators,
 $$
 \delta_1=
 \frac{2\,(2a^3+3a^2+a)}
 {2\,a\,(4+2a)}=
 {\frac { \left( a+1 \right)  \left( 2\,a+1 \right) }{2\,(a+2)}}
 $$
and
 $$
 \delta_2=
 \frac{2\,(2a^3+3a^2-6\,a-7)}
 {(6+2a)(2+2a)(1+2a)}
 =
 {\frac {2\,{a}^{2}+a-7}{ 2\,\left( 2\,a+1 \right)  \left( a+3 \right)
 }}\,.
 $$
The last missing element $\omega^{(N)}=\omega^{(N)}(a)$ in formula
(\ref{ans2}) is exceptional. Due to its manifest
truncation-dependence, its value must be computed,  at each  $N\geq
3$, by direct insertion in Eq.~(\ref{dieudo}). At the first few
dimensions this is an easy calculation which gives the
(incidentally, negative though comparatively simple) series of
formulae
 \be
 \omega^{(3)}= -\frac{3}{2\,(1+2\,a)}\,,\ \ \ \
 \omega^{(4)} =-{\frac {5+3\,a}{2\, \left(1+ 2\,a \right)  \left( 1+a \right)
 \left( 2+a \right) }}\,\ \ \ldots\, .
 \label{exce}
 \ee
Their extrapolation inspires the general ansatz
 \be
 \omega^{(N)}=
 -\frac{(u_{N}+v_{N}\,a)}
 {(2N-4+2a)\Gamma_{N-2}}
  \,
  \label{laterezza}
 \ee
and its subsequent confirmation giving
 \be
 u_{N}=(2\,N-3)(N-2)
 \,,\ \ \ \ \ \ \ \ \
 v_{N}=3\,N-6\,.
 \ee
This completes our closed-form construction of  pentadiagonal
solutions~(\ref{ans2}) of Dieudonn\'e's Eq.~(\ref{dieudo}) at any
matrix dimension $N=3,4,\ldots$.

\section{Discussion  \label{zacatek} }

%\subsection{Mathematics in simplified models }

In contrast to the recent theoretical experiments with discrete
models possessing point-like impurities \cite{Jin} or boundary terms
\cite{david}, the interaction in our one-parametric solvable toy
model is a smooth function of position. This is an innovation which
may be considered natural. In various limits and dynamical regimes
we may then specify energies $E_n$ and wave functions $|\psi_n\kt$
using the broad menu of formulae available for orthogonal
polynomials in question. In our paper the eigenstates of $H$ were
selected, for the sake of definiteness, in the closed form of
Gegenbauer polynomials.

In the context of mathematics the main obstacle of calling the
related solvable matrices $H$ Hamiltonians appeared in their
asymmetry (i.e., non-Hermiticity). This seemed to disqualify these
matrices from playing the role of operators of observables.
Fortunately, such a conclusion would be erroneous. The clarification
of the paradox dates back to Scholtz et al \cite{Geyer} and Bender
et al \cite{BBBJ}. We just recalled and used their argumentation in
a new concrete application.

Our method of the reconstruction of the metric based on the use of
discrete Hamiltonians and mediated by the computer-assisted solution
of Dieudonn\'e's Eq.~(\ref{dieudo}) proved very efficient. It led to
compact analytic formulae for a family of metrics. New
discrete-lattice quantum model has been found as described by the
pair of matrices $(H,\Theta)$. The first component of this pair is
the $N-$dimensional Gegenbauerian Hamiltonian $H$ which has been
chosen tridiagonal. The second component $\Theta$ of this pair is
the reconstructed (and non-unique) metric.

In a historical detour let us remind the readers that nuclear
physicists opened this Pandora's box of $\Theta \neq I$ cca twenty
years ago \cite{Geyer} when considering fermionic Hamiltonians
$\mathfrak{h}$ (acting in complicated Fock's space ${\cal H}^{(P)}$
of ``physical" states $|\psi^{(P)} \kt $) as transformed into
isospectral operators $H$ (acting in another ``friendly" space
${\cal H}^{(F)}$). The net gain was that the bound-state energies
became obtainable by the diagonalization of the simplified bosonic
Hamiltonian $H\neq H^\dagger$. The price to be paid was that the
latter operator proved {\em manifestly non-Hermitian} in the usual,
``friendly" Hilbert space ${\cal H}^{(F)}$ with trivial
$\Theta^{(F)}=I$.

In other branches of physics the recipe has been revitalized in
connection with the emergence of ${\cal PT}-$symmetric quantum
systems \cite{BG,BB,ostatni}. This opened new horizons in particle
physics \cite{Zamani} and in relativistic quantum field theory
\cite{Klevansky}. The key theoretical idea of the formalism (viz,
the nontriviality of the product $\Omega^\dagger\Omega :=\Theta \neq
I$) remained the same but the philosophy has been changed. In place
of starting from the knowledge of the physical, self-adjoint
$\mathfrak{h} = \mathfrak{h}^\dagger$ and from the subsequent {\em
clever choice} of a simplifying map $\Omega$, the updated
model-building strategy  (cf. \cite{Carl,ali}) takes a  {manifestly
non-Hermitian} ``friendly" candidate for the Hamiltonian $H
=H^{(F)}\neq \left ( H^{(F)}\right )^\dagger$ and  tries to
reconstruct the ``physical" Hamiltonian $\mathfrak{h}=H^{(P)}$ via
Eq.~(\ref{ouzo}).

Our present proposal of a new solvable model was inspired by the
main weakness of the latter scenario which lies in a huge
uncertainty and ambiguity of the assignment $H \to \mathfrak{h}$
marked, say, by an $N-$component multiindex $\lambda$ attached to
$\Omega=\Omega(\lambda)$). This ambiguity was inessential during the
nuclear-physics mappings $\mathfrak{h} \to H(\lambda)$. In ${\cal
PT}-$symmetric context and in its pseudo-Hermitian generalizations
\cite{Ali} it is more serious. It implies the {\em non-uniqueness of
physics} represented by the $\lambda-$dependent operator
$\mathfrak{h}(\lambda)$. The {\em same} initial operator $H$ admits
{\em many} experimentally non-equivalent physical interpretations.
The variations of $\lambda$ generate non-equivalent self-adjoint
Hamiltonians $\mathfrak{h}(\lambda)$. This means that the {\em same}
spectrum of energies may coexist with {\em different} observable
characteristics (e.g., coordinates \cite{fund,Batal}).

The suppression of the ambiguity of the multi-indexed mappings
$\Omega(\lambda)$ {\em and} of Hamiltonians $\mathfrak{h}(\lambda)$
may be performed, according to Scholtz et al. \cite{Geyer}, via an
explicit specification of some other observables ${\cal C}$, ${\cal
D}$ etc. They have to obey the same Dieudonn\'e's conditions of
cryptohermiticity. In practice, this goal may be achieved by
requiring that one of the observables used for this purpose is a
charge with involutivity property ${\cal C}^2=I$ \cite{Carl}. In our
present considerations we used another strategy proposed in
Ref.~\cite{fund} and based on the hypothesis of existence of a
nontrivial, fundamental ``smearing" length.

We showed in \cite{fund} that the smearing length does not vanish
and does not diverge in models where some of the the metrics possess
the $(2k+1)-$diagonal band-matrix form $\Theta=\Theta_k$. The
subscript $k=0,1,\ldots$ has been interpreted there as the measure
of the size of the smearing.

The simplest physical scenario of this form certainly emerges when
one decides to use just the diagonal metrics $\Theta_0\neq I$. In
Ref.~\cite{fund} as well as in our present concrete model this
``no-smearing" option proved allowed. The related diagonal-matrix
operator of the coordinate remained merely scaling-non-invariant.
Our quantum Hamiltonians then became tractable as living on {\em
deformed} but still local one-dimensional discrete $N-$site
lattices.

Once we turn attention to our present model and to its generic
band-matrix metrics $\Theta_k(a)$ with $1\leq k\ll N$, the picture
is changed and the coordinates  prove smeared~\cite{smeared}. This
feature could make our elementary solvable model tractable, e.g., as
a weakly and controllably non-local alternative to a deformed local
$k=0$ lattice~\cite{fund,fundgra}.

On the {\em experimental} level one expects that such a weakly
nonlocal scenario and its consequences (including, e.g., phase
transitions) might find simulations in classical systems. A decisive
theoretical as well as experimental progress in this direction has
already been reported in optics~\cite{Giu,Makris}. The practical
implementation of the parallel experimenting in quantum world is
hindered by several mutually interrelated obstacles. The most
serious one may be identified with a certain conflict between the
simplicity of the matrix $H$ and the complicated guarantee of its
Hermiticity via metric $\Theta$. Our resolution of this conflict has
been based on the {\em simultaneous} simplicity of {\em both} the
operators $H$ and $\Theta$.

The main {\em theoretical} profit provided by the fully
non-numerical tractability of our model may be seen in its manifest
compatibility with postulates of Quantum Mechanics in which one
works, simultaneously \cite{SIGMA}, with a {\em triplet} of
Hilbert-space representations ${\cal H}^{(P,F,S)}$ of the quantum
system in question. The Hermiticity status of operators depends on
the space but they only stay non-Hermitian in the ``naive" and
``false" space ${\cal H}^{(F)}$. Thus, in our model, the knowledge
of the friendly input matrix $H \neq H^\dagger$ is complemented by
the equally friendly nature of the  {\em ad hoc} metric
$\Theta=\Theta^{(S)}\neq I$ and, {\it ipso facto}, of the
reconstructed standard Hilbert space ${\cal H}^{(S)}$.

In practical terms our Gegenbauerian example exhibits several
specific friendly features. First of all, it is nontrivial that our
metrics are banded. This property only followed from the explicit
solution of the Dieudonn\'e's equation. Secondly, the matrix
elements of the pseudo-metrics (i.e., of the sparse-matrix
components ${\cal P}_j$ of the metrics) emerged as elementary
functions of the free real parameter $a$. Last but not least, the
matrix elements of the diagonal, tridiagonal and pentadiagonal
metrics exhibited even an almost complete independence of the
truncation~$N$.

All of these features of our Gegenbauerian model reconfirm the
feasibility of our original intention of finding a new
model-building recipe. Certainly, this (and similar) solvable models
would guarantee a viability of fitting many {\em measured} (and not
just equidistant) $N-$plets of levels $E_n^{(experimental)}$  by the
suitable $N-$plets $E_n^{(theoretical)}$ of the well known zeros of
an appropriate (i.e., in our exemplification, Gegenbauer) classical
orthogonal polynomial.

%\vspace{15mm}

\section*{Acknowledgement}

Work supported by the M\v{S}MT ``Doppler Institute" project Nr.
LC06002 and by the Institutional Research Plan AV0Z10480505.

%\newpage

%\section*{Figure captions}

%
%
%\subsection*{Figure \ref{firmone}. The boundary curve $\gamma=\gamma(\varepsilon)$.}
%
%\subsection*{Figure \ref{fione}. Spectrum of $H^{(4)}(\lambda)$.}
%
%\subsection*{Figure \ref{fitwo}. Spectrum of $H^{(6)}(\lambda)$.}
%

\newpage
\section*{Appendix A. Long-range metrics with $k=N-1$}

Equations (\ref{ans2}) + (\ref{exce}) with $N=3$ offer the simplest
nontrivial example of the metric $\Theta_k^{(N)}$ with maximal
$k=N-1$ in which some of the matrix elements become
truncation-dependent. We found that this form of manifest
$N-$dependence characterizes all the Gegenbauer metrics with $k\geq
2$. In this sense, the diagonal and tridiagonal metrics appear
exceptional. In principle, one could hope that a similar
exceptionality could characterize the antidiagonal-like metrics
which were found in some other models \cite{fund} and which could
tentatively be characterized by the triangularity property
 \be
 \left [{\cal P}_{N-1}^{(N)}(a)\right ]_{jk}=0 \ \ \ {\rm for} \ \ \
 j<k\,.
 \ee
The failure of these expectations can already be detected at the
next dimension $N=4$ because the explicit violation of
antidiagonality already characterizes the heptadiagonal pseudometric
 \be
 {\cal P}_3^{(4)}(a)=
 \left[ \begin {array}{cccc} 0&0&0&a
 \\\noalign{\medskip}0&0&{\frac
  {{a}^{2}+2\,a+1}{a+3}}&0
 \\\noalign{\medskip}0&{\frac
 {{a}^{2}+2\,a+1}{a+3 }}&0&-{\frac {3\,a+5}{ \left( a+3 \right)
 \left( 2\,a+1 \right) }}
 \\\noalign{\medskip}a&0&-{\frac {3\,a+5}{ \left( a+3 \right)  \left( 2
 \,a+1 \right) }}&0\end {array} \right]\,.
 \label{34}
 \ee
We see that the  loss of the up-down symmetry is transferred from
the Hamiltonian $H$ to the metric. Thus, one can only expect that at
a given $N$, the most elementary longest-range component ${\cal
P}_{N-1}^{(N)}(a)$ of the Gegenbauer metrics will possess the
following triangular equal-color form
 \be
 {\cal P}_{N-1}^{(N)}(a)=
 \left[ \begin {array}{ccccccc}
 0&0&0&\ldots&0&0&p_{11}
 \\
 0&0&\ldots&0&0&p_{12}&0
 \\
 \vdots&\vdots&_{\Large {\bf _.}}.^{\large \bf\, .}&
 _{\Large {\bf _.}}.^{\large \bf\, .}
 &_{\Large {\bf _.}}.^{\large \bf\, .}&0&p_{21}
 \\
 \vdots&0&0
 &p_{14}&
 _{\Large {\bf _.}}.^{\large \bf\, .}&_{\Large {\bf _.}}.^{\large \bf\, .}&0
 \\
 %\vdots&&_{\Large {\bf _.}}.^{\large \bf\, .}
% &_{\Large {\bf _.}}.^{\large \bf\, .}&
% _{\Large {\bf _.}}.^{\large \bf\, .}&
% _{\Large {\bf _.}}.^{\large \bf\, .}
% &_{\Large {\bf _.}}.^{\large \bf\, .}&
% p_{33}
% \\
  0&0&p_{13}&0&p_{23}&
  _{\Large {\bf _.}}.^{\large \bf\, .}
  &_{\Large {\bf _.}}.^{\large \bf\, .}
  \\
  0&p_{12}&0&p_{22}&0&p_{32}
  &\ldots
  \\
 p_{11}&0&p_{21}&0&p_{31}&0&\ldots
 \end {array} \right]\,.
 \label{ansNt}
 \ee
The quick growth of complexity of the, presumably, closed but much
less compact formulae for the matrix elements in (\ref{ansNt}) may
be illustrated for intermediate $N=8$ for which the maximal-range
15-diagonal (pseudo)metric matrix may be constructed by solving
Eq.~(\ref{dieudo}) via ansatz (\ref{ansNt}). In the normalization
where $\left [{\cal P}_{N-1}^{(N)}(a)\right ]_{1N}=a$ our
calculations yielded the elements
 \ben
 p_{11}=a=\frac{2\,a^2+a}{
2\,a+1
  }\,,\ \ \ \ \ p_{12}={\frac
 { \left( a+1 \right)  \left( a+3 \right) }{a+7}}
={\frac
 { 2\,{a}^{3}+9\,{a}^{2}+10\,a+3}{\left(
2\,a+1
 \right) \left (a+7 \right)}}\,,
  \een
  \ben
 p_{13}={\frac {2\,{a}^{4}+17\,{a}^{3}+52\,{a}^{2}+67\,a+30}{ \left(
 2\,a+1
 \right)  \left( a+7 \right)  \left( a+6 \right) }}=
 {\frac {\left( a+1 \right)  \left( a+3 \right)  \left( 2\,a+5 \right)  \left( a+2 \right)
 }{ \left( 2\,a+1
 \right)  \left( a+7 \right)  \left( a+6 \right) }}\,,
  \een
  \ben
 p_{14}={\frac
 {2\,{a}^{5}+25\,{a}^{4}+124\,{ a}^{3}+305\,{a}^{2}+372\,a+180}{
 \left( 2\,a+1 \right)  \left( a+7
 \right)  \left( a+6 \right)  \left( a+5 \right) }}=
 \een
 \ben
 ={\frac
 {\left( 2\,a+5 \right)  \left( a+3 \right) ^{2} \left( a+2 \right)
 ^{2} }{ \left( 2\,a+1 \right)  \left( a+7
 \right)  \left( a+6 \right)  \left( a+5 \right) }}
 \,,
  \een
plus perceivably less compact
%\newpage
  \ben
 p_{21}=-3\,{\frac
{3\,{a }^{2}+19\,a+26}{ \left( 2\,a+1 \right)  \left( a+7 \right)
\left( a+6
 \right) }}=-3\,{\frac
{\left( 3\,a+13 \right)  \left( a+2 \right) }{ \left( 2\,a+1 \right)
\left( a+7 \right) \left( a+6
 \right) }}\,,
  \een
  \ben
 p_{22}=-5\,{\frac {3\,{a}^{5}+52\,{a}^{4}
 +342\,{a}^{3}+1064\,{a}^{2}+
 1551\,a+828
 }{ \left( 2\, a+1 \right)  \left( a+6 \right)  \left( a+5 \right)
\left( a+1
 \right)  \left( a+7 \right) ^{2}}}=
  \een
 \ben
 =-5\,{\frac {\left( a+4 \right)  \left( 3\,{a}^{2}+22\,a+23 \right)  \left( a+3 \right) ^{2}
 }{ \left( 2\, a+1 \right)  \left( a+6 \right)  \left( a+5 \right)
\left( a+1
 \right)  \left( a+7 \right) ^{2}}}\,,
  \een
  \ben
 p_{23}=-6\,{\frac {6\,{
a}^{7}+143\,{a}^{6}+\ldots +40218\, {a}^{2}+37901\,a+14640}{ \left(
a+5 \right)  \left( 2\,a+3 \right)
 \left( 2\,a+1 \right)  \left( a+1 \right)  \left( a+7 \right) ^{2}
 \left( a+6 \right) ^{2}}}=
  \een
 \ben
 =-6\,{\frac {\left( 2\,a+5 \right)  \left( a+3 \right)
  \left( 3\,{a}^{5}+55\,{a}^{4}+380\,{a}^{3}+1223\,{a}^{2}+1811\,a+976 \right)
 }{ \left( a+5 \right)  \left( 2\,a+3 \right)
 \left( 2\,a+1 \right)  \left( a+1 \right)  \left( a+7 \right) ^{2}
 \left( a+6 \right) ^{2}}}\,,
  \een
(where the higher-degree polynomials in numerators have non-integer
roots which are all real) and
%\newpage
  \ben
 p_{31}=-2\,{\frac {\left( a+4 \right)
  \left( 3\,{a}^{5}+25\,{a}^{4}-78\,{a}^{3}
  -\ldots -2025 \right)}{ \left( a+5 \right)  \left( 2\,a+3
 \right)  \left( 2\,a+1 \right)  \left( a+1 \right)  \left( a+7
 \right) ^{2} \left( a+6 \right) ^{2}}}
 \,,
  \een
(where the fifth-degree polynomial in numerator has solely three
real non-integer roots),
   \ben
 p_{32}=-3\,{\frac {\left( a+3 \right)  \left( 6\,{a}^{7}+95\,{a}^{6}
 +176\,{a}^{5}-5106\,{a}^{4}-\ldots -82280 \right)
}{2\, \left( a+5 \right)  \left( 2\,a+3
 \right)  \left( 2\,a+1 \right)  \left( a+2 \right)  \left( a+1
 \right)  \left( a+6 \right) ^{2} \left( a+7 \right) ^{3}}}\,,
  \een
(where the seventh-degree polynomial in numerator has solely five
real non-integer roots) and, finally,
%\newpage
  \ben
 p_{41}=-{\frac {6\,{a}^{8}+59
\,{a}^{7}-621\,{a}^{6}-\ldots -60120}{ 2\,\left( 2\,a+5 \right)
\left( 2\,a+3
 \right)  \left( 2\,a+1 \right)  \left( a+5 \right)  \left( a+2
 \right)  \left( a+1 \right)  \left( a+6 \right) ^{2} \left( a+7
 \right) ^{3}}}\,
 \
 \een
(with just four real and four complex roots of the eighth-degree
polynomial in the numerator). Summarizing, these results demonstrate
not only the efficiency of our computer-assisted algorithms but
also, in parallel, the quick decrease of the {\em practical} appeal
of working with more-than-pentadiagonal metrics $\Theta_k^{(N)}(a)$
with $k\gg 2$.

\end{document}